# Influence of pulsed magnetic field on single- and two-pulse nuclear spin echoes in multidomain magnets.

**A. Akhkalkatsi, T. Gegechkori[a], G. Mamniashvili[a], Z. Shermadini[a,]*, A.N. Pogorely[b], O.M. Kuzmak[b]**

I.Javakhishvili Tbilisi State University,
[a]E.Andronikashvili Institute of Physics.
[b] Institute for Magnetism, National Academy of Science of Ukraine

*Corresponding author.
E-mail address: mamni@iphac.ge
(G.Mamniashvili)

## ABSTRACT

By the method of additional pulsed magnetic field influence in different magnetic materials (half metals, manganites, lithium ferrite, cobalt) it is established the analogy of time diagrams of magnetic pulse influence on single- and two pulse echoes in magnets when the distortion mechanism of single-pulse echo formation is effective, and the absence of such analogy in the case of lithium ferrite where the multipulse mechanism of single-pulse echo formation is effective.

It is shown also that the timing and frequency diagrams of magnetic pulse influence on the two-pulse echo signals , corresponding to the symmetric and asymmetric magnetic pulse applications in the studied magnets, are defined by their domain walls parameters and could serve for their qualitative and quantitative characterization.

Investigation of nuclear spin echo in magnetically ordered crystals at excitation by additional pulsed magnetic field of comparatively low (as compared with anisotropy fields) intensity turned out to be an effective tool to observe effects connected with a hyperfine field (HF) anisotropy, carry out quantitative estimations of local inhomogeneities and domain walls (DW) mobility, what is interesting for optimization of operation of different type of magnetoelectronic devices [1].

So, in work [2] the influence of pulsed magnetic field on $Eu^{151}$ nuclear spin echo signals of nuclei, arranged in DW of europium garnet $Eu_3Fe_5O_{12}$, was explained for the first time by HF field anisotropy. In work [3] it was considered the influence of pulsed magnetic field on nuclear spin echo in DW of spinel ferrites and spin echo signals in thin magnetic films.

It was studied two cases of magnetic pulse arrangement – the symmetric and asymmetric ones relative to the second radio-frequency (RF) pulse in the two-pulse method. At symmetric arrangement the decrease of echo intensity in samples with a DW originated NMR is explained by consecutive excitation of nuclei stepwisely changing their location in DW, but at asymmetric one – by the inhomogeneous shifts of NMR local frequencies due to the HF field anisotropy. It was presented experimental dependences of echo intensities on amplitude and, correspondingly, on duration of magnetic pulse.

In addition, it was shown that, while recording spin-echo signal intensity dependence on a magnetic pulse amplitude $H_d$, one could define the value of magnetic field amplitude shifting DW on its width.

Besides it, using the echo-signal intensity dependence at asymmetric excitation by magnetic pulse on a magnetic pulse duration one could calculate the density of inhomogeneous shifts distribution of nuclear frequencies caused by inhomogeneous component of HF field.

Frequency measurements demonstrated a non-uniform degree of influence of an asymmetric magnetic pulse on the echo in different parts of the resonance line of $^{59}$Co in cobalt thin magnetic films [3]. The influence was weakest in the fcc phase range (212 MHz) and strongest in the hcc phase range (218 MHz). This could serve as additional characteristics of investigated magnets, as example, in respect of stacking faults or impurities influence. The frequency spectra of the symmetric pulse influence could also turn out to be useful for the investigation of NMR line nature in magnets.

These results were found out their development in work [4] where it was investigated the influence of asymmetric magnetic field pulses (dephasing) on spin echoes of $^{59}$Co nuclei in $Y_2Co_{17}$ and $(Y_{0.9}Gd_{0.1})_2Co_{17}$. It was revealed the influence of Gd ions on the anisotropic component of the HF field in a narrow frequency band of a wide NMR spectrum. The difference between dephasing spectra in $Y_2Co_{17}$ and $(Y_{0.9}Gd_{0.1})_2Co_{17}$ shows that we have the useful method to study the changes of magnetic properties of $Y_2Co_{17}$ during the substitution by Gd. Most interesting peculiarity of dephasing method is a possibility to resolve of specific behavior of Gd ions in a very narrow frequency range what could be related with one Co-position. Really, the fact that Gd ions stipulate such strong change of anisotropic component of HF filed on $^{59}$Co nuclei in a very narrow frequency band makes it possible to conclude that they prefer some definite crystallographic positions. This way one could achieve larger resolution of NMR spectra.

For practical applications it is also perspective the effect of formation of additional echo signals at influence of magnetic pulse applied in combination with a RF pulse [5,6].

As a matter of fact the single-pulse echo (SPE) – a resonance response of nuclear spin-system on a single radio-frequency (RF) pulse, in a number of magnets behaves mainly similarly to the usual two-pulse echo (TPE).

For investigations of the nature of this analogy in the work [7] it was studied the role of RF pulse front distortions in the formation of SPE signal, when distorted edges of RF pulse play role of two RF pulses in the Hahn mechanism of two-pulse echo formation.

To clarify the role of pulse distortions in the formation of SPE it was used a technique of suppressing of the influence of the RF pulse edges on the nuclear spin system. This was achieved by applying a dc magnetic field pulse of width $\tau_d$ and amplitude $H_d$ whose timing could be varied in respect to RF pulse. Earlier, this type of excitation has been used in a two-pulse echo experiment to estimate the static field required to shift a domain wall through a distance equal to its thickness d [8].

In Fig.1 it is presented Fig.4 from [7] to illustrate the experimental layout and results obtained in work [7].

As it is seen from Fig.1, the influence of magnetic pulse for the both excitation cases of echo signals is similar. On this basis in the cited work it was made conclusion that the SPE signal in the investigated sample (lithium ferrite) is formed by the distortion mechanism when the distorted RF pulse edges play a role of RF pulses in the two-pulse method.

Further on, in the work [7] it was carried out more detailed analysis of magnetic pulse influence of the TPE intensity. As it has already been noted, the main effect of applied magnetic pulse to multidomain magnets is the displacement of DW which is reversible at small pulse amplitudes. Therefore, if the magnetic pulse is superposed on one of the RF pulses, it changes the location of the resonating nuclei in the DW (y-direction) with respect to its center thereby reducing the nuclear enhancement factors for 180$^{o}$C Bloch walls accordingly the known dependence $\eta = \eta_o \text{sech}(y/d)$, where $\eta_o$ is the maximum enhancement at the center of the wall. Therefore, if in the absence of the magnetic pulse the turning angles $\alpha_i$ of two RF pulses $\alpha_{1,2} = \gamma h \eta \tau_{1,2}$ were equal to each other in order to maximize the intensity of TPE [7,9], after the application of magnetic pulse in coincidence with one of RF pulses it follows that turning angles would differ significantly what should result in a fast reduction of TPE amplitude.

Besides it, due to the dependence of nuclear resonance frequency on location in the DW in systems with anisotropic HF field [10], the application of magnetic pulse in the interval between RF pulses, or after the second pulse would also result in the decrease of TPE signal. This effect (the dephasing effect) arises because the frequency shift partially destroys the phase coherence reducing the effectiveness of the rephasing process.

On the basis of aforementioned considerations farther on in work [7] it has been made the conclusion that regardless of the temporal location of the magnetic pulse the TPE amplitude decreases and the TPE intensity decrease is much larger in the case when $H_d$ is applied coincident with one of the RF pulses.

The interest of present authors to the considered problem is connected with our own previous investigations of the SPE formation mechanism in lithium ferrite [11,12], what didn't reveal any contribution of the distortion mechanism to the SPE in this material which was formed by the multipulse mechanism. It has been shown that the SPE signal disappears in the limit of single-pulse excitation. Besides it, the controllable additional external frequency distortion of exciting RF pulse edge similarly to work [13] resulted only in the decrease of SPE signal. Along with it a similar influence considerably (several times) enhances the SPE signal in a number of other materials, e.g., Co and half-metallic $Co_2MnSi$ and NiMnSb, where, besides it, the SPE is observed also in the single-pulse excitation mode.

Further on, in work [7] at investigation of influence of the magnetic pulse timing in respect to the SPE signal, as it has already been noted, it was made conclusion that the decrease of TPE intensity would be the most strong when $H_d$ is applied coincident with RF pulses. But allowing for results of work [3] one could suppose that the influence of magnetic pulse on the TPE signal in magnetically soft lithium ferrite with a small HF field anisotropy and high mobility of DW could differ from the one for magnetically hard materials with a relatively small DW mobility and large HF field anisotropy, such, for example, as cobalt.

To illustrate these considerations it was carried out experiments to study the influence of pulsed magnetic field with amplitude up to 500 Oe and duration of several $\mu$s using NMR spectrometer, a pulsed magnetic field source, and samples described in details in work [7,5], correspondingly. Measurements were made on cobalt, lithium ferrite, half metals, such as NiMnSb, $Co_2MnSi$ and manganites, attracting currently great interest from the point of view of their applications in spintronics [14], and on cobalt thin magnetic films.

The obtained results confirm the main conclusion of work [7] that there is an analogy in the influence of magnetic pulse on SPE and TPE in systems where the echo signal is formed by the distortion mechanism. Let us present for illustration the corresponding dependences for cobalt and NiMnSb in Fig. 2-3. But in the case of lithium ferrite, when the echo signal is formed by the multipulse mechanism, the magnetic pulse influence picture on the SPE signal is different, Fig.4.

In this case the MP influence is similar to the MP influence on three-pulse stimulated echo when the MP is applied between the first and the second RF pulses, Fig.5. The observed influence picture corresponds to the stimulated character of the SPE formation in lithium ferrite [12,15].

As it has already been pointed out, frequency diagrams of the symmetrical MP influence could be useful, along with ones of asymmetric MP application [3,4] for the characterization of the investigated magnets. Let us show this an example of polycrystal cobalt and lithium ferrite, Fig, 6 and 7, correspondingly, and for half metal NiMnSb, Fig.8.

In the case of half metallic $Co_2MnSi$ it was received the frequency diagrams of MP influence for $^{55}Mn$ and $^{59}Co$ nuclei, showing the significant difference in the degree of local HF fields anisotropy of corresponding nuclei, Fig.9. For visualization, let us present here as well time diagrams of MP influence for both cases, Fig. 10.

As per TPE excitation, the picture of pulse influence in polycrystal Co and thin Co films qualitatively differs from that one observed in materials with a relatively large DW mobility and a small HF field anisotropy and it is also different from the dependence assumed in [7]. Therefore, time diagrams of magnetic pulsed field influence on TPE could be used for the characterization of magnetic materials. For illustration, let us compare the time diagrams of MP influence on TPE signals in cobalt, Fig.2a, with that of NiMnSb, Fig. 3a, for which there is an analogy with lithium ferrite, Fig.4a.

In addition, the frequency diagrams of symmetric pulse influence reflect the peculiar physical properties of materials and could be used for their characterization as ones for asymmetric pulse [3,4].

In conclusion, it is established by the method of magnetic pulse influence the analogy between the timing diagrams of magnetic pulse influence of TPE and SPE of magnets where SPE is formed by the distortion mechanism, in difference with the case of lithium ferrite where SPE is formed by multipulse mechanism.

It was shown also that the magnetic pulse influence timing and frequency diagrams could be used for additional sensitive characterization of magnetic materials in both cases of symmetric and asymmetric excitations along with usual NMR spectra.

This work was supported by the STCU Grant Ge-051 (J).

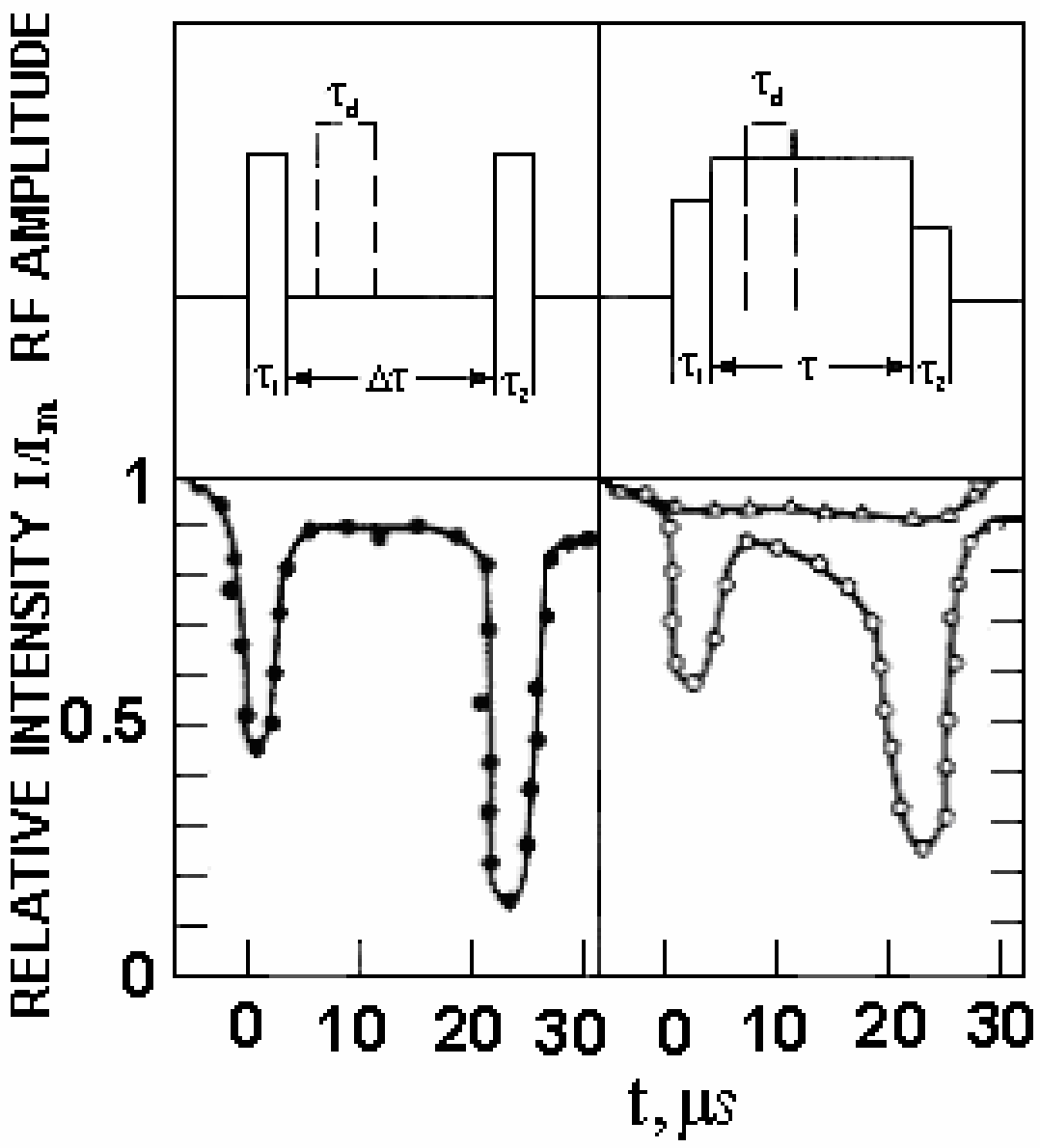

Fig. 1.

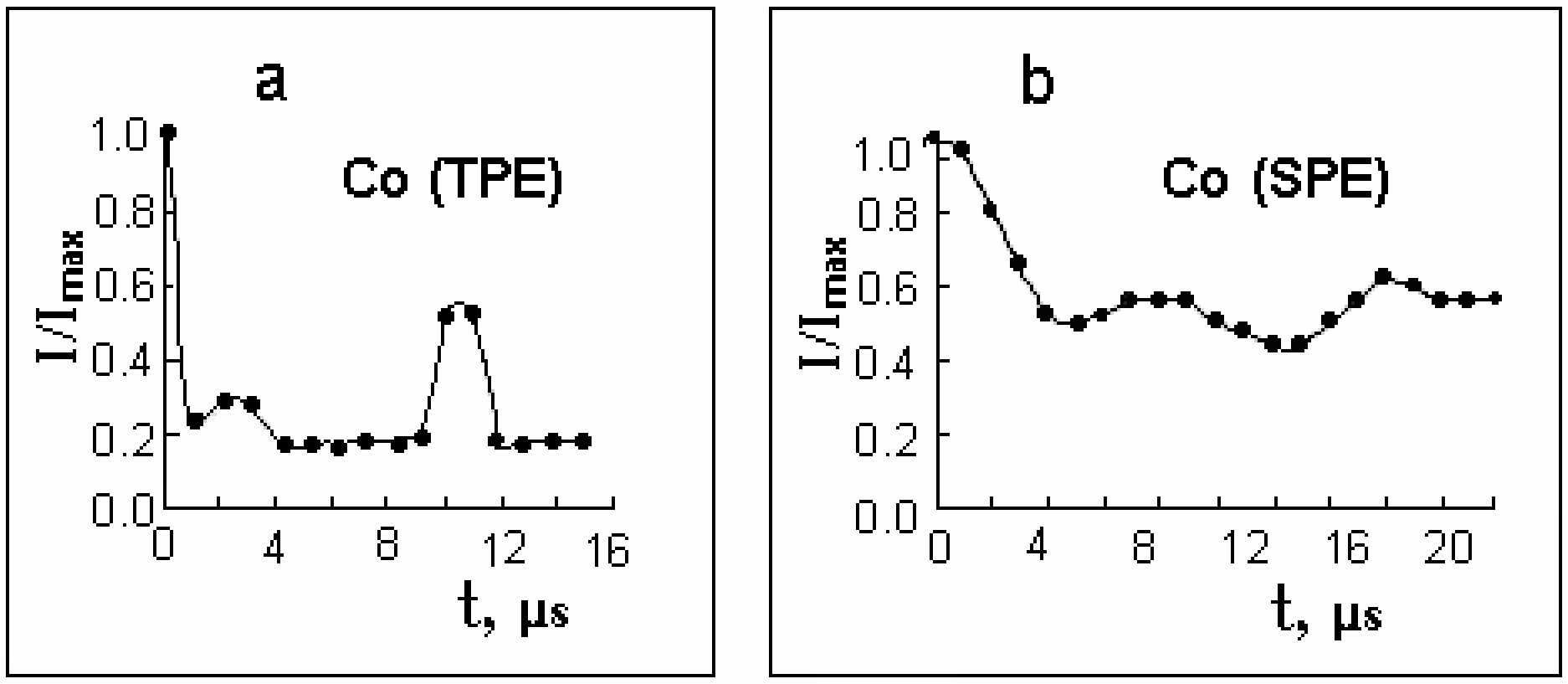

Fig. 2.

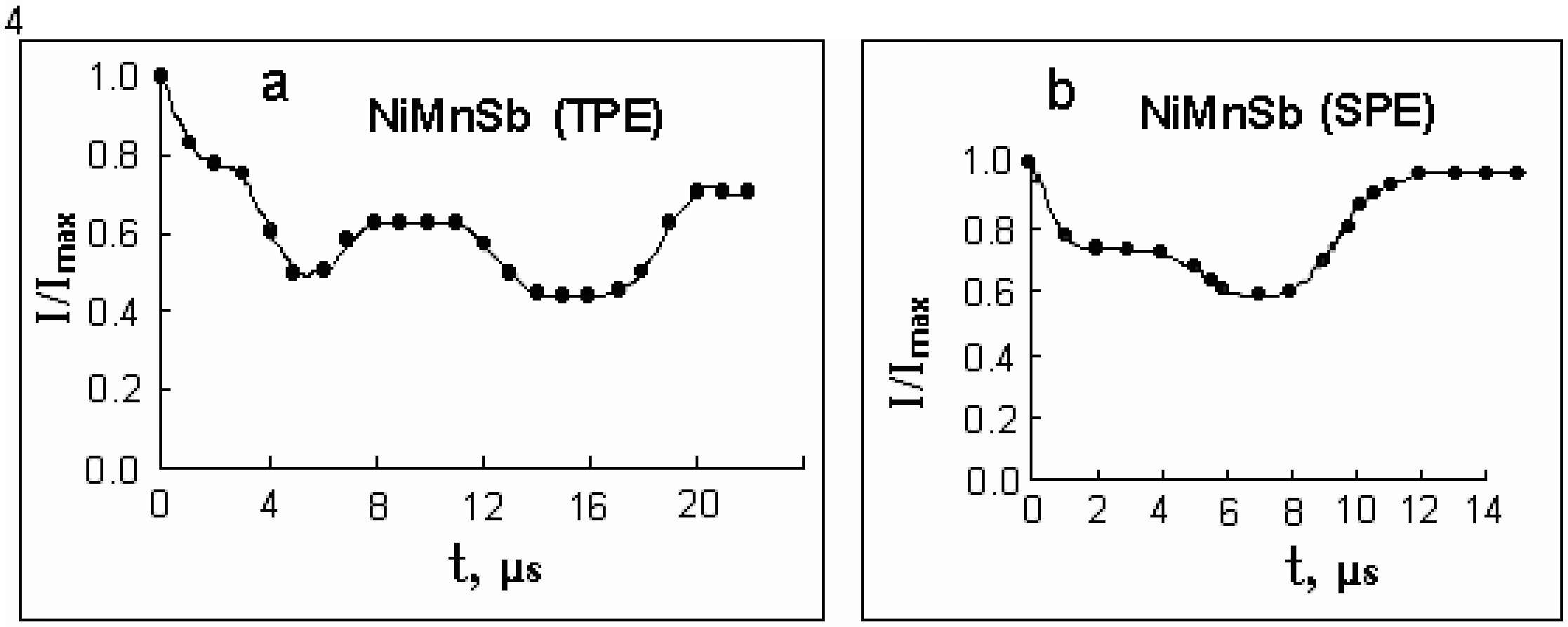

Fig. 3.

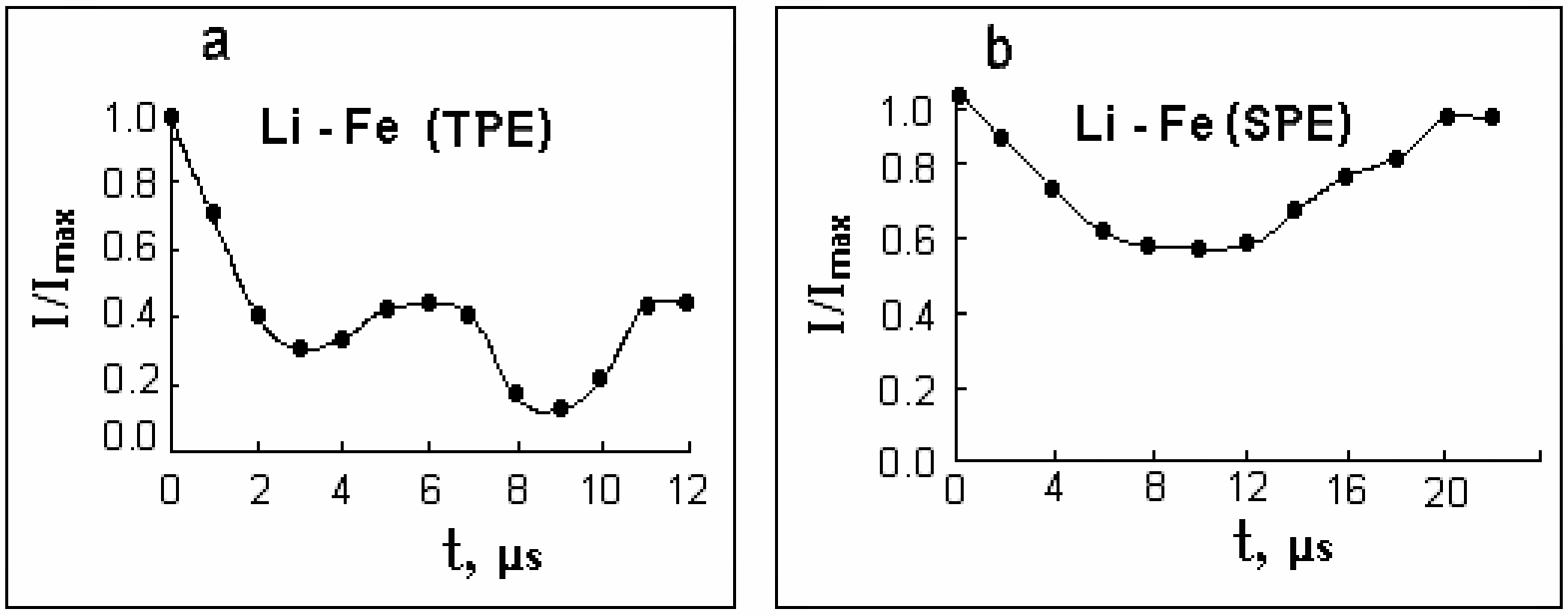
a
Li - Fe (TPE)
I/I_max
t, μs
b
Li - Fe (SPE)
I/I_max
t, μs

Fig. 4.

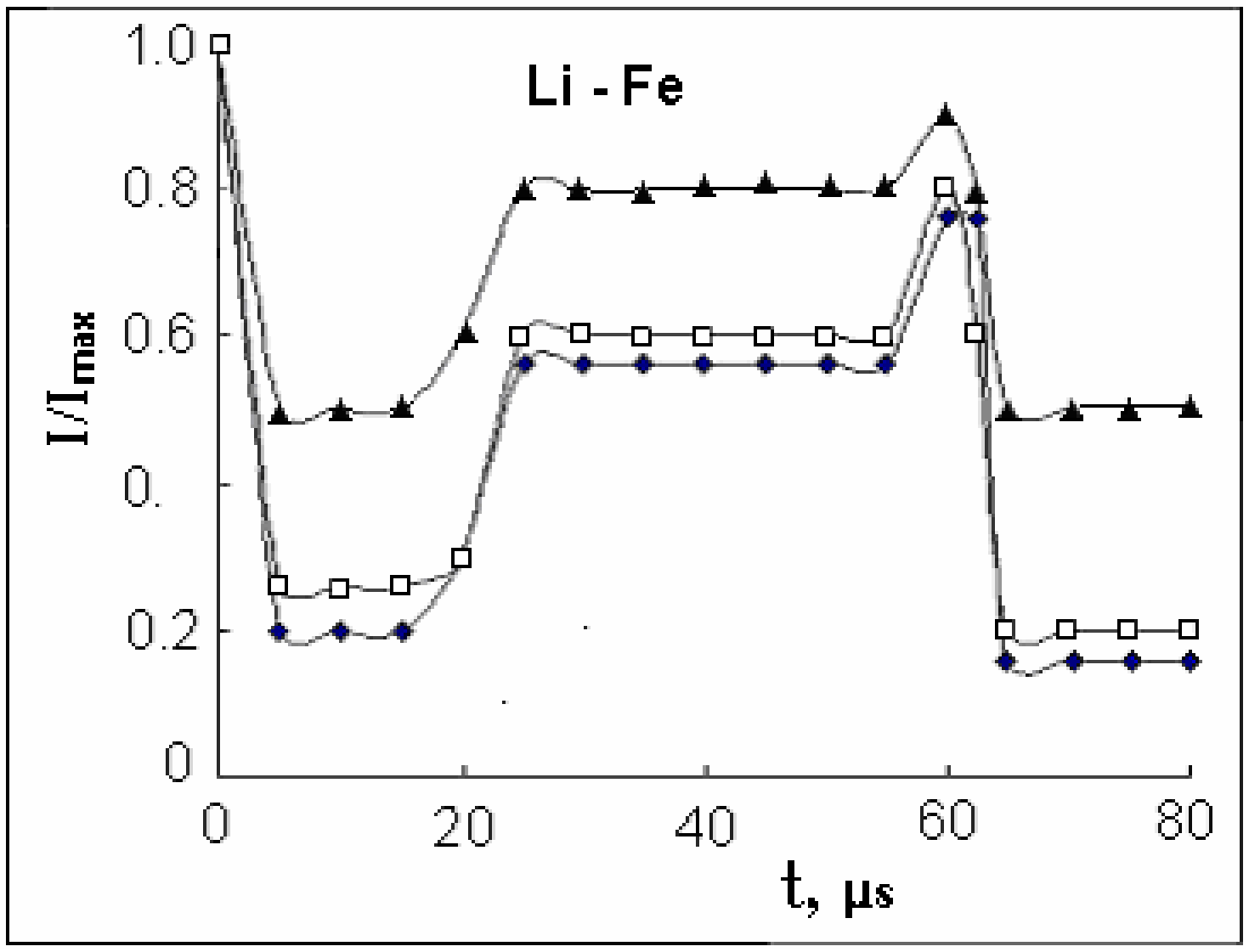
Li - Fe
I/I_max
t, μs

Fig. 5.

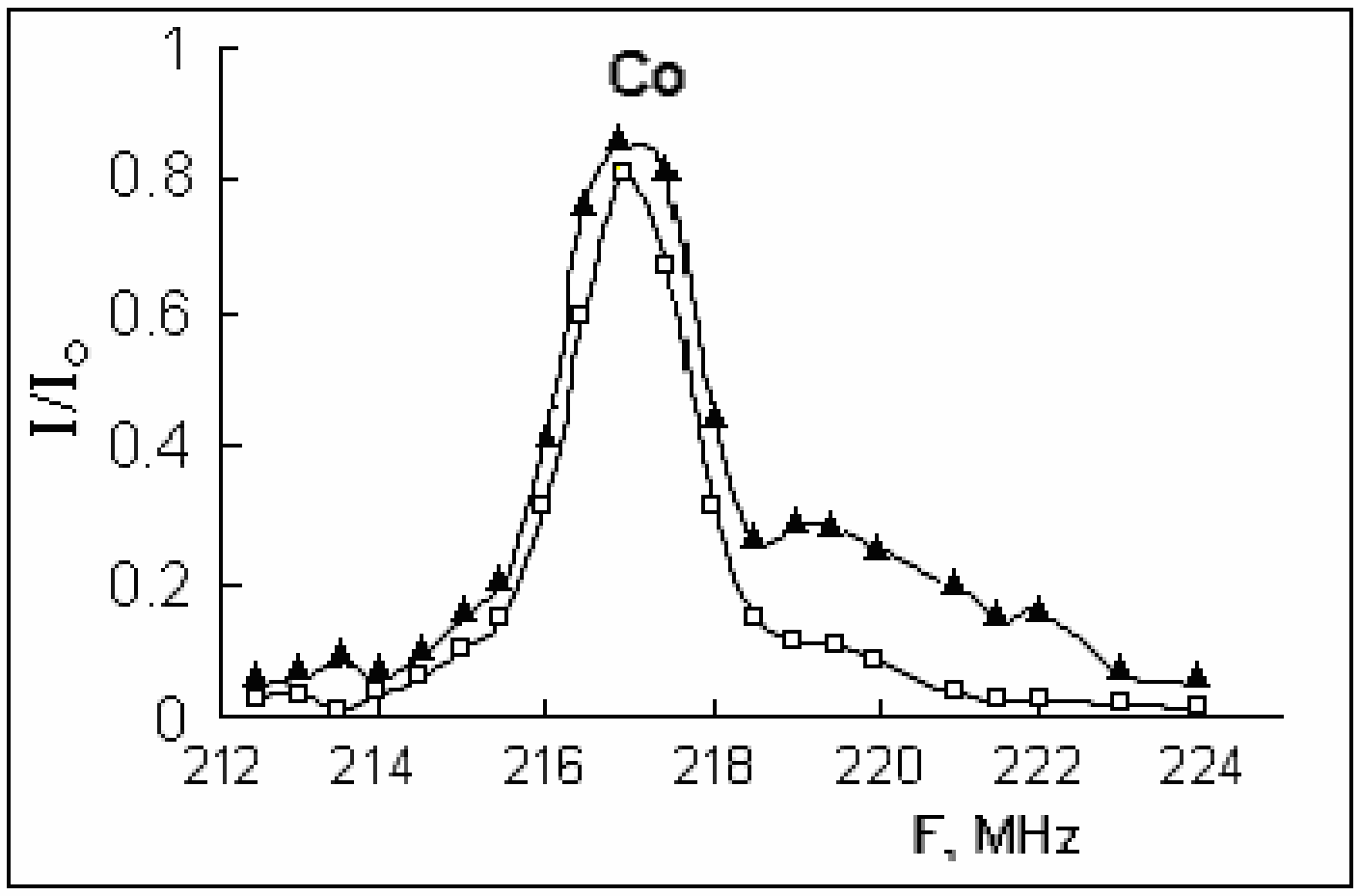
Co
I/I_o
F, MHz

Fig. 6.

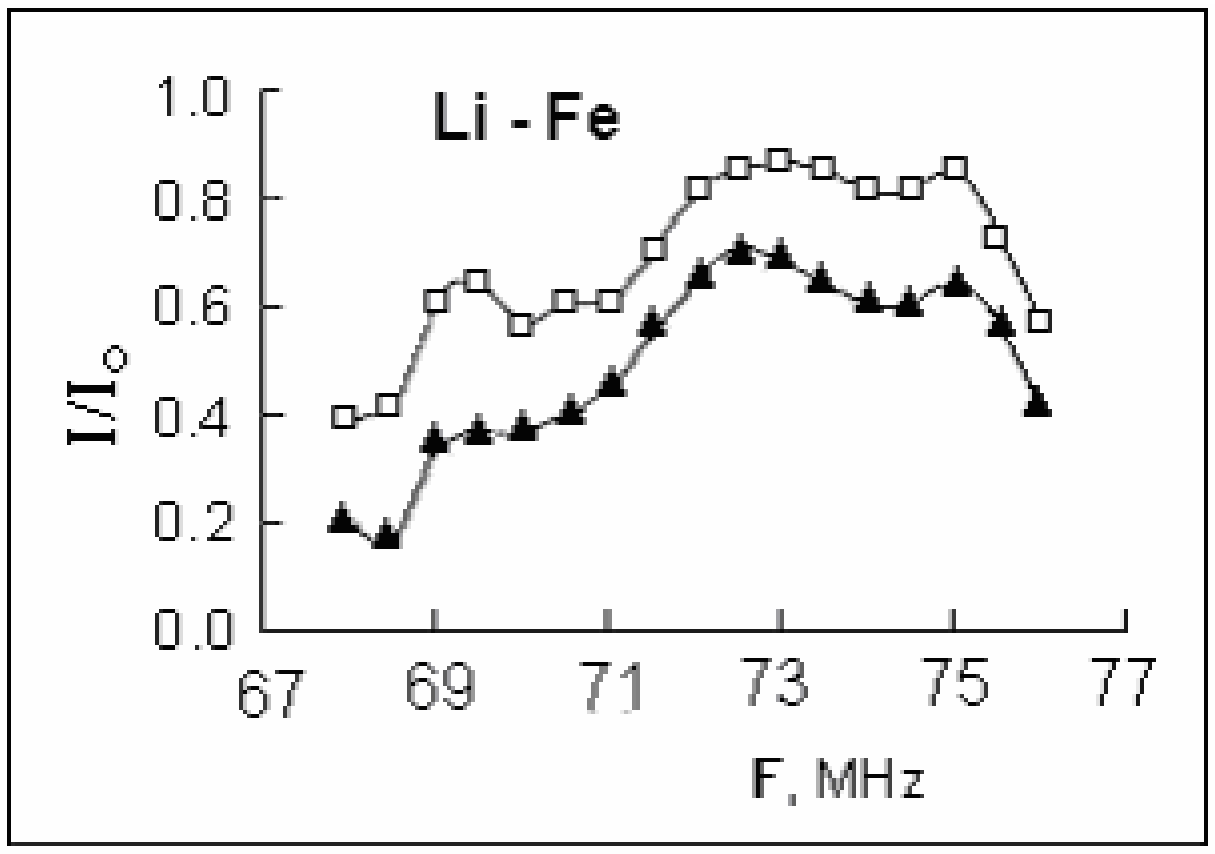

Fig. 7.

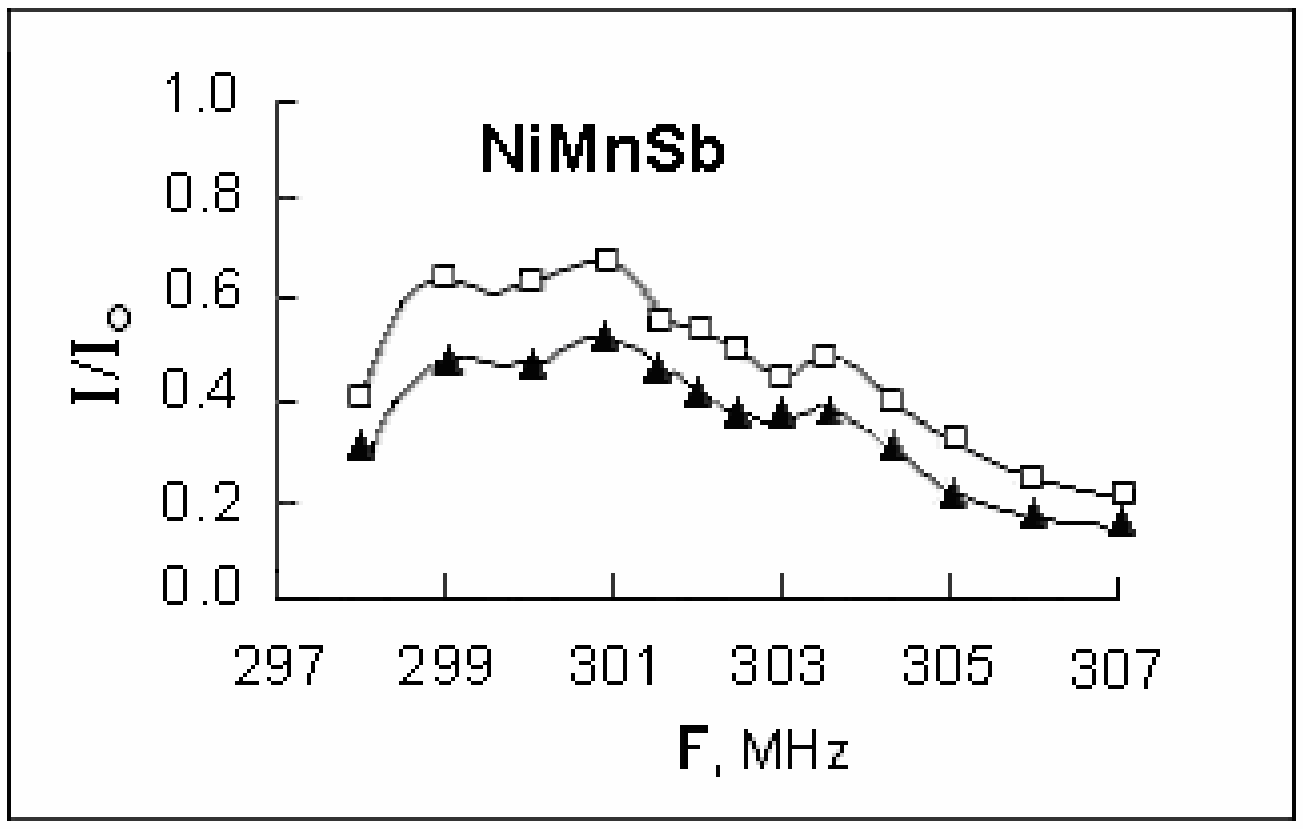

Fig. 8.

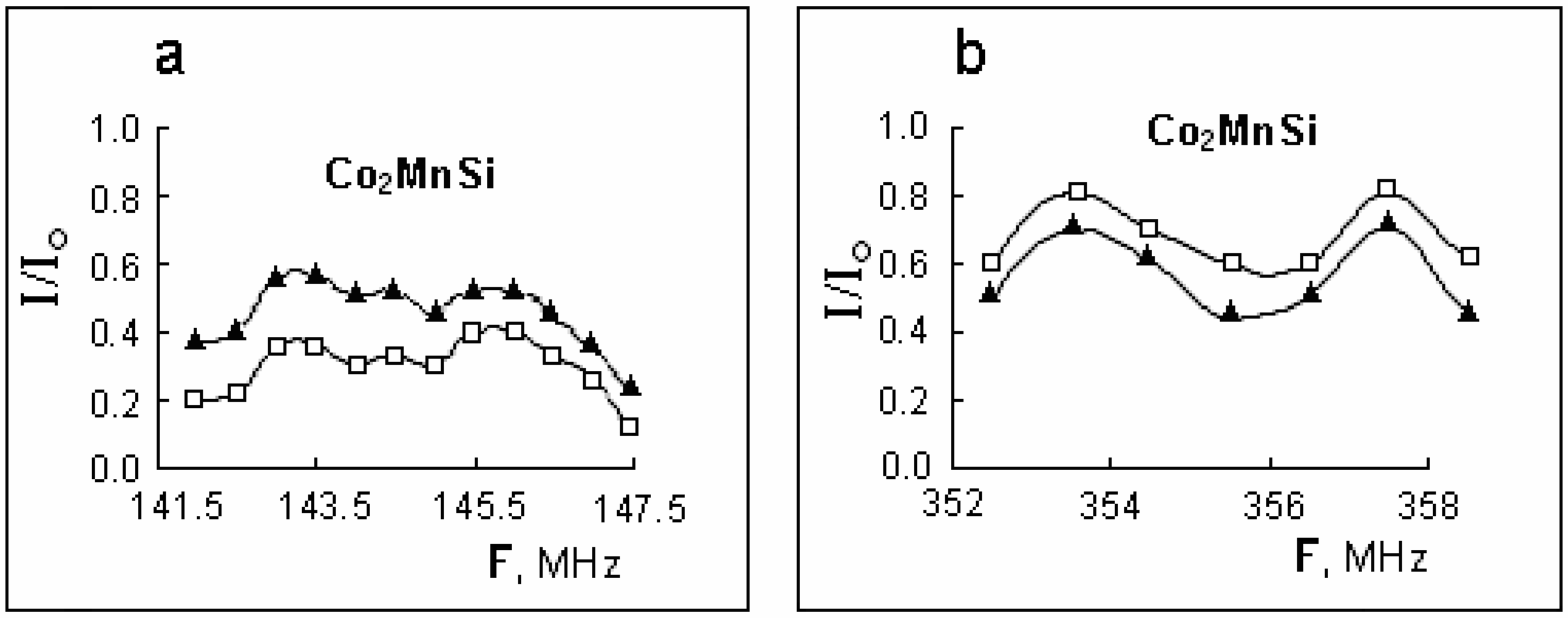

Fig. 9.

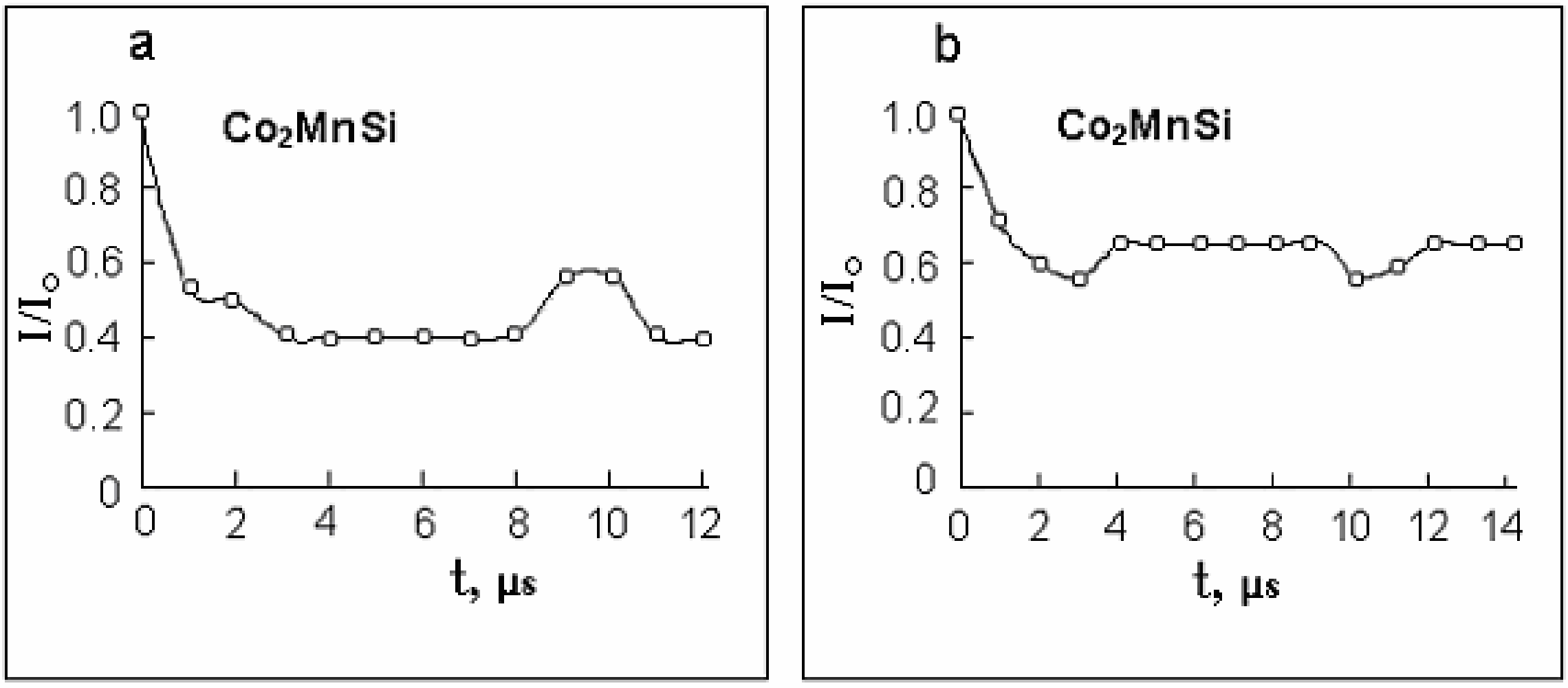

Fig. 10.

# FIGURE CAPTIONS

Fig. 1. Timing diagrams of the intensity dependence of the two-pulse echo(●), single-pulse echo (○) and FID's original maximum (Δ) on the temporal location of a dc magnetic pulse of $H_d$=5 Oe lasting for a time interval $\tau_d$ along the corresponding time scales for the $^{57}$Fe NMR in lithium ferrite. The data for two-pulse echo were taken with $\tau_1 = \tau_2 = 0.8$ μs, $\Delta\tau = 21$ μs, $\tau_d = 3$ μs, while the single-pulse echo data correspond to $\tau_1 = 25$ μs, $\tau_d = 5$ μs. Resonance frequency 74.0 MHz, T=77 K.

Fig. 2. Timing diagrams of the intensity dependence of the two-pulse echo (a) and single-pulse echo (b) on the temporal location of the $H_d$ magnetic pulse with duration $\tau_d$ in polycrystal cobalt at:

a) $\tau_1 = \tau_2 = 1.6$ μs, $\Delta\tau = 10$ μs, $\tau_d = 2.4$ μs, $H_d = 100$ Oe, $f_{NMR} = 216.5$ MHz;

b) $\tau = 17$ μs, $\tau_d = 1$ μs, $H_d$=30 Oe, $f_{NMR}$=216.5 MHz.

Fig. 3. Timing diagrams of the intensity dependence of the two-pulse echo (a) and single-pulse echo (b) on the temporal location of the $H_d$ magnetic pulse with duration $\tau_d$ in half metal NiMnSb at:

a) $\tau_1 = \tau_2 = 2$ μs, $\Delta\tau = 11$ μs, $\tau_d = 3$ μs, $H_d$=150 Oe, $f_{NMR}$=300 MHz;

b) $\tau = 10$ μs, $\tau_d = 1$ μs, $H_d$=30 Oe, $f_{NMR}$=300 MHz.

Fig. 4. Timing diagrams of the intensity dependence of the two-pulse echo (a) and single-pulse echo (b) on the temporal location of the $H_d$ magnetic pulse with duration $\tau_d$ in lithium ferrite ($Li_{0.5}Fe_{2.35}Zn_{0.15}O_4$) at:

a) $\tau_1 = 1$ μs, $\tau_2 = 1.4$ μs, $\Delta\tau = 7$ μs, $\tau_d = 1.7$ μs, $H_d$=28 Oe, $f_{NMR}$=74 MHz;

b) $\tau = 20$ μs, $\tau_d = 2$ μs, $H_d$=10 Oe, $f_{NMR}$=71.5 MHz.

Fig. 5 Timing diagrams of the intensity dependence of the three-pulse stimulated echo on the temporal location of the magnetic pulse in lithium ferrite ($Li_{0.5}Fe_{2.35}Zn_{0.15}O_4$) at:

$\tau_1 = 0.8$ μs, $\tau_2 = 1$ μs, $\tau_3 = 1$ μs, $\tau_d = 3$ μs, $H_d$=40 Oe, $\tau_{12} = 20$ μs, $\tau_{23} = 40$ μs,

◆ $f_{NMR}$=71 MHz, □ $f_{NMR}$=72 MHz, ▲ $f_{NMR}$=73 MHz.

Fig.6. Frequency dependence of the influence of the symmetric (▲) and asymmetric (□) magnetic pulses on two-pulse echo in cobalt at:

$\tau_1 = 1.1$ μs, $\tau_2 = 1.2$ μs, $\Delta\tau = 10$ μs, $\tau_d = 2$ μs, $H_d$=100 Oe. $I_o$ – echo amplitude at $H_d$=0.

Fig.7. Frequency dependence of the effect of the symmetric (▲) and asymmetric (□) magnetic pulses on two-pulse echo in lithium ferrite ($Li_{0.5}Fe_{2.35}Zn_{0.15}O_4$) at:

$\tau_1 = 0.8$ μs, $\tau_2 = 0.9$ μs, $\Delta\tau = 14$ μs, $\tau_d = 1$ μs, $H_d$=15 Oe; $I_o$ – echo amplitude at $H_d$=0.

Fig.8. Frequency dependence of the effect of the symmetric (▲) and asymmetric (□) magnetic pulses on two-pulse echo in NiMnSb at:

$\tau_1 = \tau_2 = 2$ μs, $\Delta\tau = 10$ μs, $\tau_d = 3$ μs, $H_d$=150 Oe; $I_o$ – echo amplitude at $H_d$=0.

Fig. 9. Frequency dependence of the effect of the symmetric (▲) and asymmetric (□) magnetic

pulses on two-pulse echoes in $Co_2MnSi$ for $^{59}Co$ NMR (a) and $^{55}Mn$ NMR(b) at:
a) $\tau_1$ = 1.1 μs, $\tau_2$ = 1.4 μs, $\Delta\tau$ = 10 μs, $\tau_d$ = 2 μs, $H_d$=550 Oe ($^{59}Co$ ЯМР);
b) $\tau_1$ = 0.8 μs, $\tau_2$ = 1 μs, $\Delta\tau$ = 13 μs, $\tau_d$ = 4 μs, $H_d$=190 Oe ($^{55}Mn$ ЯМР);
$I_o$ – echo amplitude at $H_d$=0.

Рис.10. Timing diagrams and intensity dependence of the two-pulse echo on the temporal location of the magnetic pulse with duration $\tau_d$ in $Co_2MnSi$ for $^{59}Co$ NMR (a) and $^{55}Mn$ NMR (b) at:
a) $\tau_1$ = 1.1 μs, $\tau_2$ = 1.4 μs, $\Delta\tau$ = 10 μs, $\tau_d$ = 2 μs, f=145,5 MHz, $H_d$=550 Oe;
b) $\tau_1$ = $\tau_2$ = 3 μs, $\Delta\tau$ = 7 μs, $\tau_d$ = 2 μs, $H_d$=300 Oe, f=354 MHz.